\documentclass[conference]{IEEEtran}
\IEEEoverridecommandlockouts 
\usepackage[letterpaper, top=0.75in, bottom=1.2in, left=0.68in, right=0.68in]{geometry}

\usepackage{cite}
\usepackage{amsmath,amssymb,amsfonts}
\usepackage{algorithmic}
\usepackage{graphicx}
\usepackage{textcomp}
\usepackage{xcolor}
\usepackage{adjustbox}
\usepackage{tikz}
\usepackage{booktabs}
\usepackage{array}
\usepackage{svg}
\usepackage{subcaption}
\usepackage[numbers]{natbib}

\usetikzlibrary{shapes,arrows}
\usepackage{pdflscape}
\def\BibTeX{{\rm B\kern-.05em{\sc i\kern-.025em b}\kern-.08em
    T\kern-.1667em\lower.7ex\hbox{E}\kern-.125emX}}
\usepackage{array}
\usepackage{colortbl}

\begin{document}
\title{A Multitask VAE for Time Series Preprocessing and Prediction of Blood Glucose Level
}

\author{\IEEEauthorblockN{Ali ABUSALEH}
\IEEEauthorblockA{
\textit{Air Liquide R\&D}\\
Les Loges-en-Josas 78350, France \\
}
\and
\IEEEauthorblockN{Mehdi RAHIM}
\IEEEauthorblockA{
\textit{Air Liquide R\&D}\\
Les Loges-en-Josas 78350, France \\
mehdi.rahim@airliquide.com}

}

\maketitle

\begin{abstract}

Data preprocessing is a critical part of time series data analysis. Data from connected medical devices often have missing or abnormal values during acquisition. Handling such situations requires additional assumptions and domain knowledge.
This can be time-consuming, and can introduce a significant bias affecting predictive model accuracy and thus, medical interpretation.
To overcome this issue, we propose a new deep learning model to mitigate the preprocessing assumptions.
The model architecture relies on a variational auto-encoder (VAE) to produce a preprocessing latent space, and a recurrent VAE to preserve the temporal dynamics of the data.
We demonstrate the effectiveness of such an architecture on telemonitoring data to forecast glucose-level of diabetic patients. Our results show an improvement in terms of accuracy with respect of existing state-of-the-art methods and architectures.
\end{abstract}

\begin{IEEEkeywords}
Generative Deep Learning, Data Preprocessing, Time Series, VAE. 
\end{IEEEkeywords}

\section{Introduction}

Diabetes mellitus (DM) is a complex and pervasive metabolic disorder characterized by strong variations of the blood glucose (BG) level. The physiopathology is complex and mainly caused by defects in insulin secretion, insulin action, or both.
There are several types of diabetes, including type 1 diabetes mellitus (T1DM) where the patient has a complete lack of insulin production, type 2 diabetes mellitus (T2DM) where the body's cells become less responsive to insulin, and relative insulin deficiency, where the pancreas fails to produce enough insulin to compensate for insulin resistance, and other less common forms.

Glucose-level forecasting from connected continuous glucose monitoring (CGM) devices help anticipate hyperglycmia (high glucose level) or hypoglycemia (low glucose level). CGM can measure glucose level every five minutes, enabling an \textit{almost} real time remote monitoring of the patient health status.  
The taxonomy of blood glucose level forecasting can be categorized into two main approaches \cite{diabetesSurvay}: \textit{i) Physiological models}, which requires an in-depth understanding of an individual's physiological mechanisms. This approach involves the study of how the human metabolism works and how it affects blood glucose levels. This often referred to as \textit{white-box} \cite{diabetesBoxes}; \textit{ii) Data-driven models}, which utilizes historical data as a foundation for predictions. This approach is often referred to as a \textit{black-box} \cite{diabetesBoxes} model. It does not require an understanding of the underlying physiological mechanisms. However, a comprehensive understanding of the data and proper preprocessing are crucial.
In particular, data preprocessing has a significant impact on the quality of the data-driven forecasting model of glucose-level. Yet, there are no standard preprocessing pipelines for CGM readings.

To mitigate the preprocessing assumptions, we propose a new deep learning model.
The model architecture relies on a variational auto-encoder (VAE) to produce a preprocessing latent space, and a recurrent VAE to preserve the temporal dynamics of the data.
We apply this model in the context of blood glucose forecast for patients with Diabetes. Our results highlight the effectiveness of our approach compared to baselines and state-of-the art predictive models.
Our technical \textbf{contributions} are two-fold :
\begin{itemize}
    \item \textbf{Architecture}: A novel VAE architecture that incorporates a temporal attention mechanism. This architecture is a self-contained preprocessing model and a prediction network.
    \item \textbf{Loss}: Optimized loss functions tailored for enhancing forecasting accuracy in healthcare applications.
\end{itemize}
The paper is organized as follows : section \ref{sec:context} introduces the methods related to time series forecasting with deep learning models. Section \ref{sec:arch} describe our contribution and architecture. Section \ref{sec:experiment} shows the experiments and results. Section \ref{sec:conclusion} concludes the paper and discusses future works.

\section{Deep Learning for Time Series}
\label{sec:context}

Time series are a widespread data modality to characterize phenomena in numerous domains such as finance, healthcare, and industry. 
Proper handling and preprocessing of these data are crucial to ensure accurate and reliable analyses and predictions. Preprocessing techniques for time series data encompass different items such as missing values, outliers, and feature scaling. If uncorrected, time series analysis can be significantly distorted.
Missing values in time series data are often due to sensor malfunctions or data transmission errors. Techniques such as Multiple Imputation (MI), are effective in addressing these issues \cite{tawakuli2024}. Similarly, outlier detection methods, help identify and mitigate the impact of anomalous data points that could skew results \cite{tawakuli2024}, normalization, Feature extraction and selection, and many other factors play a pivotal role in the model outcome.
Time series preprocessing includes techniques such as normalization (\cite{uvaPaduva}, \cite{neuroFuzzy}), scaling \cite{uvaPaduva}, smoothing \cite{RnnPrediction}, approximation, interpolations. Each step has an impact on the final representation of the time series.
Generative models, such as Variational Autoencoders (VAEs), have shown promise in representing data in a latent space \cite{kingma2022autoencoding}. While traditional VAEs have demonstrated their effectiveness in representation learning, they suffer from limitations in handling time series data. Recent advances have proposed novel VAE architectures to address these limitations. For example, \cite{vaeTimeSeries1} proposed a VAE with a recurrent neural network (RNN) encoder and decoder, while \cite{vaeTimeSeries2} proposed a VAE with a temporal convolutional network (TCN) encoder and decoder. These architectures have shown improved performance in representing time-series data compared to traditional VAEs.
In the context of blood glucose level forecasting, VAEs have shown potential for handling missing data and generating synthetic data. For instance, \cite{vaeBg1} use a VAE to impute missing clinical data in electronic health records, while \cite{vaeBg2} use a VAE to generate synthetic blood glucose data for a population of patients.
However, blood glucose level forecasting presents unique challenges that traditional VAEs may not fully address. Factors such as nutrition and physical exercises, and the irregular usage  of CGM devices, complicate the forecasting task. 
To overcome this, we propose a novel VAE architecture that incorporates a temporal attention mechanism. This enables the VAE to better capture the dynamics, performing a preprocessing and improve forecasting accuracy.
We introduce in the following section our contribution. To the best of our knowledge, this model is an original architecture tailored for both time series preprocessing and forecasting.

\section{Model Architecture}
\label{sec:arch}
We describe in this section our model to \textbf{reduce the dependency on preprocessing techniques} while maintaining the main objective of \textbf{long-term forecasting and accuracy}. This model combines different architectures as discussed in \cite{vaeBg1} and \cite{vaeTimeSeries1}.
Our proposed model leverages the Variational Autoencoder (VAE) framework to encode time series data into a latent space. Additionally, we integrate Recurrent Neural Networks (RNNs) within the VAE architecture to preserve the temporal dependencies inherent in the sequential data. This hybrid approach not only ensures robust data imputation but also enhances the accuracy of long-term forecasting. It captures both individual data point nuances and overall temporal patterns.
We detail below the main components of the model 

\subsection{Data in the latent space}

\paragraph*{Mathematical Formulation}
The VAE represents the data generation process using a probabilistic framework. The encoder maps the input \( \mathbf{x} \) to a probability distribution over the latent space, \( q_\phi(\mathbf{z}|\mathbf{x}) \), where \( \phi \) are the parameters of the encoder. This distribution is typically chosen to be a Gaussian distribution with mean \( \mu \) and variance \( \sigma^2 \).

\begin{equation}
q_\phi(\mathbf{z}|\mathbf{x}) = \mathcal{N}(\mathbf{z}; \mu(\mathbf{x}), \sigma^2(\mathbf{x}))    
\end{equation}

The decoder then maps the latent variable \( \mathbf{z} \) back to a distribution over the input space, \( p_\theta(\mathbf{x}|\mathbf{z}) \), where \( \theta \) are the parameters of the decoder. The reconstruction of the input is

\begin{equation}
p_\theta(\mathbf{x}|\mathbf{z}) = \mathcal{N}(\mathbf{x}; \hat{\mathbf{x}}, \sigma_x^2)    
\end{equation}

To train the VAE, we maximize the evidence lower bound (ELBO) on the marginal likelihood of the data. The ELBO consists of two terms: a reconstruction loss and a regularization term. The reconstruction loss measures how well the decoder can reconstruct the input data, while the regularization term enforces a prior distribution \( p(\mathbf{z}) \) (typically a standard normal distribution) on the latent variables.
\begin{equation}
\mathcal{L}(\phi, \theta; \mathbf{x}) = \mathbb{E}_{q_\phi(\mathbf{z}|\mathbf{x})}[\log p_\theta(\mathbf{x}|\mathbf{z})] - D_{\text{KL}}(q_\phi(\mathbf{z}|\mathbf{x}) || p(\mathbf{z}))    
\end{equation}
Here, \( \mathbb{E}_{q_\phi(\mathbf{z}|\mathbf{x})}[\log p_\theta(\mathbf{x}|\mathbf{z})] \) is the expected reconstruction loss and \( D_{\text{KL}} \) is the Kullback-Leibler divergence, which measures the difference between the learned latent distribution \( q_\phi(\mathbf{z}|\mathbf{x}) \) and the prior distribution \( p(\mathbf{z}) \).

\subsection{Temporal Dynamics in Latent Space}

While a standard VAE offers advantages in the latent space representation, it may not effectively capture the temporal dependencies inherent in sequential data like time series. To address this limitation, we incorporate an RNN as an encoder-decoder component of the VAE \cite{vaeTimeSeries1}. This modification aims to preserve the temporal dynamics of the data, allowing the model to better capture the sequential dependencies present in time series data.

The Variational Recurrent Neural Network (VRNN) \cite{vaeTimeSeries1} integrates a VAE at each time step $t$ conditioned on the previous RNN state $h_{t-1}$. This configuration enables the VAE to account for the sequential nature of the data more effectively than a traditional VAE. Unlike a standard VAE, the latent variable prior in VRNN follows a non-standard Gaussian distribution, and the generating distribution is conditioned not only on $z_t$ but also on $h_{t-1}$. The RNN updates its hidden state $h_t$ using a recurrence equation that incorporates $\phi_{x}^{\tau}$, $\phi_{z}^{\tau}$, and $h_{t-1}$. This setup defines the distributions $p(z_t | x_{<t}, z_{<t})$ and $p(x_t | z_{\leq t}, x_{<t})$. Similarly, the approximate posterior $q(z_t | x_{\leq t}, z_{<t})$ is conditioned on $x_t$ and $h_{t-1}$. The encoding of the posterior and decoding for generation are linked through $h_{t-1}$, resulting in the factorization \cite{vaeTimeSeries1}
\begin{equation}
    q(z_{\leq T} | x_{\leq T}) = \prod_{t=1}^{T} q(z_t | x_{\leq t}, z_{<t})
\end{equation}
\subsection{A synthetic example}
\begin{figure}[htbp]
  \centering
  \begin{subfigure}{\linewidth}
    \centering
    \includegraphics[width=\linewidth]{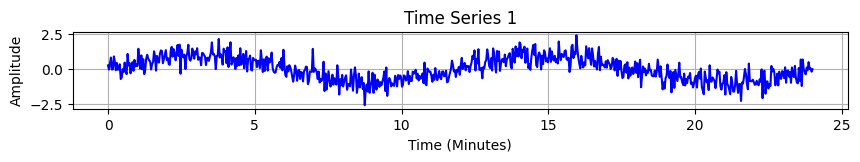}
    \text{a- Original Time Series $ts_1$}
  \end{subfigure}\\
  \begin{subfigure}{\linewidth}
    \centering
    \includegraphics[width=\linewidth]{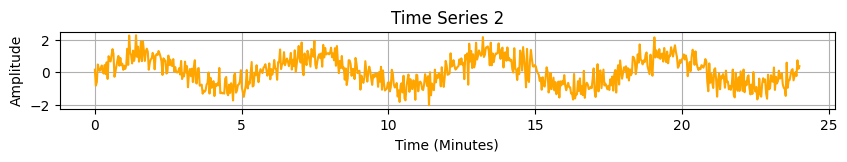}
    \text{b- Original Time Series $ts_2$}
  \end{subfigure}\\
  \begin{subfigure}{\linewidth}
    \centering
    \includegraphics[width=\linewidth]{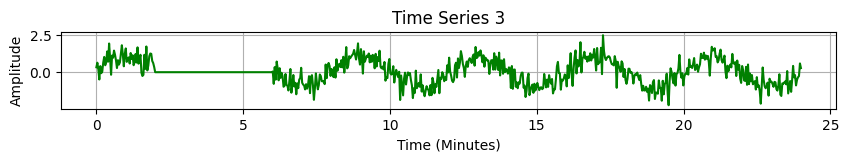}
    \text{c- Original Time Series $ts_3$}
  \end{subfigure}
  \caption{Synthetic time series to highlight the VAE based imputation.}
  \label{latent-space:original}
\end{figure}

The latent space in a VAE \cite{vaeBg1} can be used to better capture the data dynamics while preserving the individuality of each data point. The following example shows three different time series explained as follows:
\newpage
\begin{itemize}
    \item $ts_1 = \sin(2 \pi \cdot time / 12) + 0.5 \cdot \mathcal{N}(n_{\text{samples}}, 1)$
    \item $ts_2 = \sin(2 \pi \cdot time / 6) + 0.5 \cdot \mathcal{N}(n_{\text{samples}}, 1)$
    \item $ts_3 = \sin(2 \pi \cdot time / 4) + 0.5 \cdot \mathcal{N}(n_{\text{samples}}, 1)$
\end{itemize}
where $ts_1$ and $ts_2$ retain all the data points, $ts_3$ is masked for an interval of 1 hour (60 data points) to simulate a missing data scenario. Figure~\ref{latent-space:original} shows the visual representation of the three time series.
After projecting the time series into a latent space using the VAE encoder, even with missing data points, the model can learn the relationships between existing data and use them to infer the missing information. Here's the equation for projecting a time series data point ($x$) into the latent space using the VAE encoder: $z = \text{encoder}(x)$.
This equation represents encoding the data point $x$ using the encoder function of the VAE to obtain its latent representation $z$. By analyzing the relationships between latent representations in the latent space, the model can potentially estimate missing values for incomplete time series.

The masked time series ($ts_3$) is affected by the dynamics of $ts_1$ and $ts_2$, and a simulation of contextual data imputation is performed on $ts_3$. The following figure \ref{latent-space:after} shows the imputation in $ts_3$.

\begin{figure}[htbp]
  \centering
  \includegraphics[width=\linewidth]{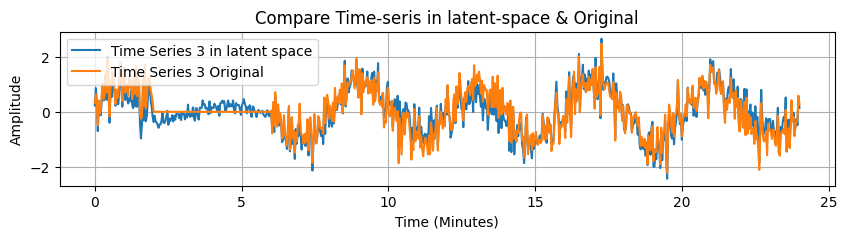}
  \caption{Comparison of $ts_3$ before and after projection to latent space}
  \label{latent-space:after}
\end{figure}

This approach offers a promising solution for handling missing data in time series forecasting by reducing reliance on manual intervention and potentially improving the accuracy of long-term forecasts. However, further research is needed to explore the effectiveness of different imputation techniques within the latent space and their

\subsection{Proposed hybrid ML Model/Architecture}
We propose a tailored version of VAE, where we have incomplete data scarcity.
We selected this architecture component after benchmarking different RNN components. This benchmark is discussed in \ref{sec:experiment}.

\subsubsection*{Architecture}
The architecture of the VAE involves an RNN-based encoder and decoder. The encoder maps the input \( \mathbf{x} \) to a latent space, represented by the mean (\(\mu\)) and the logarithm of the variance (\(\log \sigma^2\)). The latent variable \( \mathbf{z} \) is then sampled and passed through the decoder, which reconstructs the input features and predicts future glucose levels. This architecture also functions as a data imputation model, where the RNN components capture the temporal dynamics, and the latent space captures the correlations in the data, effectively handling missing periods.
The components are as follows:
\begin{itemize}
    \item An RNN encoder with input size \( d \) and hidden size \( h \).
    \item Fully connected layers to compute the mean (\(\mu\)) and log variance (\(\log \sigma^2\)).
    \item An RNN decoder to reconstruct the input and predict future values.
    \item Separate output layers for reconstruction and prediction.
\end{itemize}

\begin{figure*}[htbp]
\centering
  \centering
  \includegraphics[width=.9\linewidth]{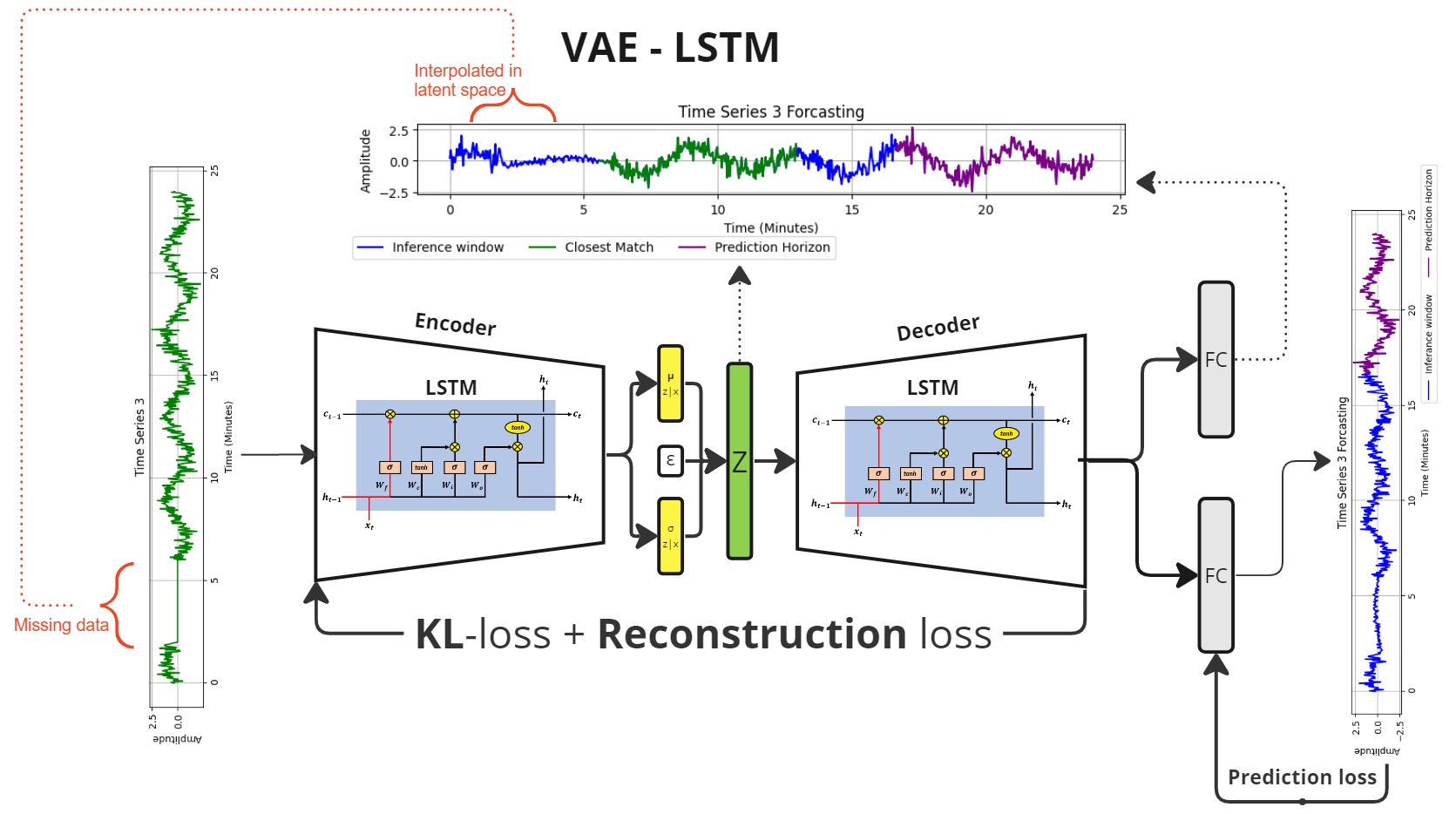}
  \caption{\textbf{Patient-specific VAE-LSTM architecture.}
The architecture consists of a fully connected layer, followed by two main components: an encoder and a decoder, both built based on GRU/LSTM. The encoder maps the input data $\mathbf{x}$ to a latent representation $\mathbf{z}$, while the decoder reconstructs the best representative of the data from this latent representation. The missing data is represented and interpolated by the latent space.}
  \label{fig:vaevariationlstm2}
\end{figure*}

 The figure~\ref{fig:vaevariationlstm2} shows an LSTM version of the proposed architecture.

\subsubsection*{Objective}
The model aims to learn a latent space representation that encapsulates the seasonal patterns, like patient's metabolism. The model also performs data preprocessing such as imputation, etc, and forecasting within a time horizon. This objective is achieved by maximizing the likelihood of the target next inference window $w$, given the input inference window \( \mathbf{x} \).
Table~\ref{tab:symbols} summarizes the notations used to describe the optimization.

\begin{table}[htbp]
\centering
\caption{Notations used and their meanings}
\begin{tabular}{ll}
\textbf{Symbol} & \textbf{Meaning} \\
\hline
\( \mathbf{x} \) & Input features \\
\( \mathbf{y} \) & Target prediction horizon \\
\( d \) & Number of features \\
\( T \) & Length of the time series \\
\( w \) & Length of the prediction horizon \\
\( \boldsymbol{\theta} \) & Parameters of the VAE \\
\( q_{\boldsymbol{\theta}}(\mathbf{z} \mid \mathbf{x}, h_{t-1}) \) & Encoder of the VAE \\
\( p_{\boldsymbol{\theta}}(\mathbf{x} \mid \mathbf{z}, h_{t-1}) \) & Decoder of the VAE \\
\( \mathbf{z} \) & Latent variable of dimension \( k \) \\
\( p(\mathbf{z}) \) & Prior distribution of the latent variable \\
\( \text{ELBO}(\boldsymbol{\theta}) \) & Evidence Lower Bound (ELBO) objective \\
\( \text{KL}(q_{\boldsymbol{\theta}}(\mathbf{z} \mid \mathbf{x}, h_{t-1}) \| p(\mathbf{z})) \) & Kullback-Leibler (KL) divergence \\
\end{tabular}
\label{tab:symbols}
\end{table}
%
%
The optimization process aims to maximize the evidence lower bound (ELBO) objective: 
\hfill
\begin{align}
\text{ELBO}(\boldsymbol{\theta}) =  & \mathbb{E}_{q_{\boldsymbol{\theta}}(\mathbf{z} \mid \mathbf{x}, h_{t-1})} \left[ \log p_{\boldsymbol{\theta}}(\mathbf{x} \mid \mathbf{z}, h_{t-1}) \right] \nonumber \\
& - \text{KL}\left(q_{\boldsymbol{\theta}}(\mathbf{z} \mid \mathbf{x}, h_{t-1}) \| p(\mathbf{z})\right)
\end{align}
\hfill


where \( \text{KL}(q_{\boldsymbol{\theta}}(\mathbf{z} \mid \mathbf{x}, h_{t-1}) \| p(\mathbf{z})) \) denotes the Kullback-Leibler (KL) divergence between the encoder's posterior distribution and the prior distribution \( p(\mathbf{z}) \).

\subsubsection*{Loss Function}
The loss function for the model comprises three components:
\begin{enumerate}
    \item \textit{The reconstruction loss} $\mathcal{L}_{\text{reco}}$ quantifies the discrepancy between the reconstructed features and the actual input features, using Mean Squared Error (MSE).
    \begin{equation}
    \mathcal{L}_{\text{reco}}
= \frac{1}{n} \sum_{i=1}^{n} \|\mathbf{x}_i - \mathbb{E}_{q_{\boldsymbol{\theta}}(\mathbf{z} \mid \mathbf{x}_i, h_{t-1})} [p_{\boldsymbol{\theta}}(\mathbf{x}_i \mid \mathbf{z}, h_{t-1})]\|^2
    \end{equation}    
    \item \textit{The prediction loss} $\mathcal{L}_{\text{pred}}$ measures the accuracy of predicted future glucose levels against actual values, using Mean Squared Error (MSE).
    \begin{equation}
        \mathcal{L}_{\text{pred}} = \frac{1}{n} \sum_{i=1}^{n} \|\mathbf{y}_i - \mathbb{E}_{q_{\boldsymbol{\theta}}(\mathbf{z} \mid \mathbf{x}_i, h_{t-1})} [p_{\boldsymbol{\theta}}(\mathbf{y}_i \mid \mathbf{z}, h_{t-1})]\|^2
    \end{equation}    
    \item \textit{The KL divergence loss} $\mathcal{L}_{\text{KL}}$ measures the divergence between the learned latent distribution and the prior distribution.
    \begin{equation}
        \mathcal{L}_{\text{KL}} = \frac{1}{n} \sum_{i=1}^{n} -0.5 \sum_j \left( 1 + \log (\sigma_j^2) - \mu_j^2 - \sigma_j^2 \right)
    \end{equation}
\end{enumerate}
\hfill

The total loss is 
{\vspace{-2.8mm}
\begin{equation}
\mathcal{L}_{\text{total}} = \alpha \mathcal{L}_{\text{reco}} + \beta \mathcal{L}_{\text{pred}}  + \gamma \mathcal{L}_{\text{KL}}.
\end{equation}}

One should note that any kind of RNN can be used, such as LSTM, GRU, etc.

\section{Experiments and Results}
\label{sec:experiment}
We present the experiments on forecasting blood glucose levels over a limited time-series dataset. 
We analyze : \textit{i)} The performance of the proposed model compared to state-of-the-art models and baselines.
We evaluate the predictive models on the public dataset OhioT1DM 2018 \cite{Marling2020-op}. The dataset contains 8 weeks of CGM, insulin, and self-reported meals for a total of 6 patients with type-1 diabetes. The dataset is primarily utilized for blood glucose prediction.

\subsection{Glucose-level prediction accuracy}
We compare our model to different RNN models (LSTM, Bi-LSTM, GRU, Bi-GRU), in addition to a statistical model ARIMA and two baselines from naive approaches: forward fill, linear trend. 
These models are trained on all patient data with an $80\%-20\%$ training-testing split and 20 epochs.

Table \ref{tab:experiment_results_30min} depicts the mean $\pm$ standard deviation of RMSE (Root Mean Squared Error), MAPE (Mean Absolute Percentage Error), and nMAPE (Normalized Mean Absolute Percentage Error) for a prediction horizon of 30 minutes (6 steps). 
Table \ref{tab:experiment_results_1h} provides the results for a prediction horizon of 1 hour (12 steps). 
Results show that Bi-GRU models while demonstrating good overall performance in terms of RMSE and MAPE, are suboptimal for our specific glucose-level prediction use case.
Conversely, our proposed VAE-GRU model (highlighted in yellow in Table \ref{tab:experiment_results_1h}) shows higher performance. It achieves the lowest RMSE and MAPE values among all models evaluated for the 1-hour prediction horizon.
\begin{table}[htbp]
    \centering
    \caption{Forecasting accuracy comparison  (30 minutes)}
    \resizebox{\columnwidth}{!}{%
    \begin{tabular}{|c|c|c|c|}
        \hline
        \textbf{Model} & \textbf{30Min\_RMSE} & \textbf{30Min\_nMAPE} & \textbf{30Min\_MAPE} \\
        \hline
        ForwardFill & 69.65 $\pm$ 28.89 & 35.04 $\pm$ 15.55 & 36.13 $\pm$ 18.55 \\
        LinearTrend & 59.88 $\pm$ 25.29 & 31.05 $\pm$ 16.03 & 31.72 $\pm$ 11.66 \\
        ARIMA & 72.88 $\pm$ 32.96 & 39.25 $\pm$ 23.29 & 38.48 $\pm$ 19.98 \\
        LSTM & 47.08 $\pm$ 19.14 & 22.93 $\pm$ 11.76 & 23.28 $\pm$ 9.37 \\
        GRU & 34.47 $\pm$ 11.87 & 16.30 $\pm$ 7.15 & 16.99 $\pm$ 7.61 \\
        BiLSTM & 33.23 $\pm$ 5.95 & 14.92 $\pm$ 3.45 & 16.12 $\pm$ 3.82 \\
        BiGRU & 30.13 $\pm$ 6.25 & 13.10 $\pm$ 3.68 & 14.01 $\pm$ 2.54 \\
        VAE-LSTM (Ours) & \cellcolor{yellow}28.26 $\pm$ 6.53 & \cellcolor{yellow}12.30 $\pm$ 3.39 & \cellcolor{yellow}12.82 $\pm$ 2.87 \\
        VAE-GRU (Ours) & 26.93 $\pm$ 7.19 & 11.73 $\pm$ 3.75 & 11.99 $\pm$ 3.11 \\
        \hline
    \end{tabular}
    }
    \label{tab:experiment_results_30min}
\end{table}
\begin{table}[htbp]
    \centering
    \caption{Forecasting accuracy comparison  (1 Hour)}
    \resizebox{\columnwidth}{!}{%
    \begin{tabular}{|c|c|c|c|}
        \hline
        \textbf{Model} &  \textbf{1H\_RMSE} & \textbf{1H\_nMAPE} & \textbf{1H\_MAPE} \\
        \hline
        ForwardFill & 68.31 $\pm$ 27.66 & 34.44 $\pm$ 14.98 & 35.73 $\pm$ 18.25 \\
        LinearTrend & 58.86 $\pm$ 24.95 & 30.61 $\pm$ 15.94 & 31.42 $\pm$ 11.55 \\
        ARIMA & 71.60 $\pm$ 31.60 & 38.72 $\pm$ 22.71 & 38.13 $\pm$ 19.67 \\
        LSTM & 51.12 $\pm$ 16.31 & 25.27 $\pm$ 10.19 & 25.62 $\pm$ 8.05 \\
        GRU & 42.18 $\pm$ 11.03 & 19.98 $\pm$ 6.75 & 21.57 $\pm$ 7.14 \\
        BiLSTM  & 43.87 $\pm$ 5.56 & 20.32 $\pm$ 3.53 & 21.35 $\pm$ 2.33 \\
        BiGRU & 40.95 $\pm$ 7.80 & 19.08 $\pm$ 4.96 & 19.87 $\pm$ 4.45 \\
        VAE-LSTM (Ours) & 40.30 $\pm$ 8.63 & 18.53 $\pm$ 4.97 & 19.46 $\pm$ 4.30 \\
        VAE-GRU (Ours) &  \cellcolor{yellow}\textbf{39.92} $\pm$ 9.73 & \cellcolor{yellow}\textbf{18.42} $\pm$ 5.64 & \cellcolor{yellow}\textbf{19.16} $\pm$ 4.59 \\
        \hline
    \end{tabular}
    }
    \label{tab:experiment_results_1h}
\end{table}
\subsubsection*{Clinical performance of glucose-level prediction}
In addition to the statistical metrics, we analyze the performance of the predictive models in terms of diabetes-related metrics called Clark Error Grid.
The Clarke Error Grid is a visual tool used to assess the accuracy of blood glucose monitoring systems or continuous glucose monitoring systems. 
It is also used to assess glucose-level forecasting.
The grid is divided into five zones: A to E. Each zone represents a different level of accuracy and potential clinical risk. Zone A is the most accurate, while zone E is an extremely large overestimation or underestimation of the true glucose value. They could lead to treatment decisions that have serious medical risk. 

As shown in Table \ref{tab:clerk_grid_error}, VAE-RNN models exhibit better clinical outcomes. Specifically, VAE-LSTM, while quantitatively superior in most metrics, shows less reliability in clinical Zone-D. Bi-RNN models have a negative impact on clinical results. Table \ref{tab:clerk_grid_error} presents the clinical metrics including A, B, C, D, and E, where higher values of A + B indicate better performance (aiming for $> 90\%$), and lower values of D + E are desired. Based on the clinical metrics presented in Table \ref{tab:clerk_grid_error}, the VAE-GRU model (highlighted in yellow) achieves the highest values for A (83.55 $\pm$ 7.64) and B (14.19 $\pm$ 6.45), indicating superior performance in predicting glucose levels above 90\%. It also shows low values for D (1.74 $\pm$ 1.26) and E (0.09 $\pm$ 0.14), demonstrating minimal errors in critical glucose level predictions, which are crucial for clinical management. In contrast, while VAE-LSTM performs well in A and B metrics, it shows higher values in D (7.54 $\pm$ 8.75) and E (0.04 $\pm$ 0.06), suggesting potential clinical limitations compared to VAE-GRU.

\begin{table}[htp]
    \centering
    \small
    \caption{Clarke Error Grid model comparison (1 Hour)}
     \resizebox{\columnwidth}{!}{%
    \begin{tabular}{|c|c|c|c|c|c|}
        \hline
        \textbf{Model} & \textbf{A (\%)} & \textbf{B (\%)} & \textbf{C (\%)} & \textbf{D (\%)} & \textbf{E (\%)} \\
        \hline
        ARIMA & 20.10 $\pm$ 23.53 & 41.75 $\pm$ 24.90 & 2.33 $\pm$ 4.72 & 17.65 $\pm$ 27.88 & 6.21 $\pm$ 15.20 \\
        BiGRU & 63.90 $\pm$ 31.57 & 20.25 $\pm$ 12.54 & 1.07 $\pm$ 2.18 & 4.34 $\pm$ 8.18 & 5.48 $\pm$ 13.81 \\
        BiLSTM & 62.12 $\pm$ 10.58 & 31.14 $\pm$ 8.33 & 0.24 $\pm$ 0.60 & 6.48 $\pm$ 5.07 & 0.02 $\pm$ 0.00 \\
        ForwardFill & 40.47 $\pm$ 18.29 & 44.38 $\pm$ 15.60 & 7.89 $\pm$ 14.24 & 7.58 $\pm$ 18.02 & 0.00 $\pm$ 0.00 \\
        GRU & 73.65 $\pm$ 12.89 & 21.40 $\pm$ 15.81 & 0.00 $\pm$ 0.00 & 4.90 $\pm$ 4.67 & 0.00 $\pm$ 0.00 \\
        LSTM & 52.50 $\pm$ 13.58 & 38.38 $\pm$ 8.84 & 0.97 $\pm$ 0.98 & 0.34 $\pm$ 4.29 & 2.52 $\pm$ 4.48 \\
        LinearTrend & 63.49 $\pm$ 39.45 & 44.30 $\pm$ 17.66 & 10.28 $\pm$ 18.64 & 0.00 $\pm$ 0.00 & 0.00 $\pm$ 0.00 \\
        \rowcolor{yellow} VAE-GRU (Ours) & \textbf{83.55} $\pm$ \textbf{7.64} & \textbf{14.19} $\pm$ \textbf{6.45} & \textbf{2.46} $\pm$ \textbf{1.04} & \textbf{1.74} $\pm$ \textbf{1.26} & \textbf{0.09} $\pm$ \textbf{0.14} \\
        \rowcolor{yellow} VAE-LSTM (Ours) & 64.57 $\pm$ 25.26 & 27.84 $\pm$ 17.04 & 0.00 $\pm$ 0.00 & 7.54 $\pm$ 8.75 & 0.04 $\pm$ 0.06 \\
        \hline
    \end{tabular}
    }
    \label{tab:clerk_grid_error}
\end{table}

\begin{figure}[htp]
  \includegraphics[width=1\columnwidth]{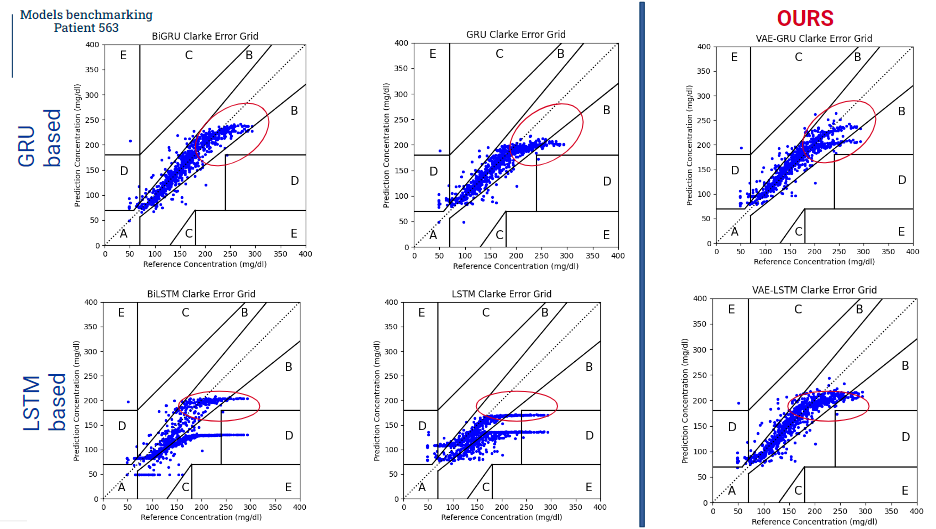}
  \caption{\textbf{Clarke Error Grid on Patient-563}. LSTM models exhibit resistance to predictions above 160, while biLSTM models have a higher threshold around 200 with adverse clinical effects. Similar observations apply to GRU models. Our proposed models demonstrate a balanced approach, achieving higher prediction thresholds while maintaining superior clinical performance.}
  \label{fig:vae_563_results}
\end{figure}

We show in figure~\ref{fig:vae_563_results} one example of Clarke Error Grid. 
The first row shows GRU models, both GRU and biGRU, while the second row displays the (Bi)-LSTM models. 
The results indicate that LSTM models exhibit resistance to predictions above 160, whereas biLSTM models have a higher threshold of around 200 with adverse clinical effects. 
Similar observations apply to GRU models. Our proposed models demonstrate a balanced approach, achieving higher prediction thresholds while maintaining superior clinical performance.

\subsubsection*{Long-term prediction horizon} We tested the proposed model in the long-term, namely 240 data points (4 hours)in the future). Table~\ref{tab:mard}  shows the Normalized Mean Absolute Percentage Error (nMAPE) values for different models
at various prediction intervals. The results for VAE-LSTM and VAE-GRU (ours) are also included for comparison with other models, and they suggest better results in the long term with better stability over a longer horizon. 

\begin{table}[htp]
\centering
\caption{Long-term prediction horizon comparison}
\resizebox{\columnwidth}{!}{%
    \begin{tabular}{|c|c|c|c|c|c|}
\toprule[0.65ex]
\textbf{Model} & \textbf{30Min nMAPE} & \textbf{1H nMAPE} & \textbf{2H nMAPE} & \textbf{3H nMAPE} & \textbf{4H nMAPE} \\
\midrule
ForwardFill & 53.5 $\pm$ 23.4 & 46.2 $\pm$ 31.7 & -- & -- & -- \\
LinearTrend & 32.8 $\pm$ 24.0 & 28.0 $\pm$ 19.7 & -- & -- & -- \\
ARIMA & 49.4 $\pm$ 17.7 & 41.2 $\pm$ 22.2 & -- & -- & -- \\
BiLSTM & 35.1 $\pm$ 20.8 & 26.3 $\pm$ 7.6 & 29.5 $\pm$ 9.1 & 29.9 $\pm$ 6.7 & 31.2 $\pm$ 7.1 \\
BiGRU & 27.1 $\pm$ 13.3 & 40.9 $\pm$ 37.6 & 44.3 $\pm$ 35.2 & 48.1 $\pm$ 36.9 & 32.6 $\pm$ 9.7 \\
LSTM & 27.6 $\pm$ 9.3 & 27.8 $\pm$ 8.7 & 32.7 $\pm$ 5.6 & 42.3 $\pm$ 32.3 & 30.5 $\pm$ 8.3 \\
GRU & 36.6 $\pm$ 13.2 & 41.3 $\pm$ 31.1 & 41.4 $\pm$ 32.9 & 35.7 $\pm$ 12.3 & 30.5 $\pm$ 8.3 \\
\rowcolor{yellow}
\textbf{VAE-LSTM (Ours)} & \textbf{23.2 $\pm$ 8.8} & \textbf{23.8 $\pm$ 9.1} & \textbf{27.4 $\pm$ 8.3} & \textbf{29.6 $\pm$ 8.2} & \textbf{30.4 $\pm$ 7.7} \\
\rowcolor{yellow}
\textbf{VAE-GRU (Ours)} & \textbf{22.1 $\pm$ 8.3} & 25.1 $\pm$ 9.6 & \textbf{27.3 $\pm$ 8.0} & 29.9 $\pm$ 8.4 & {30.7 $\pm$ 8.0} \\
\bottomrule
\end{tabular}}
\label{tab:mard}
\end{table}


\subsection{Learning speed of predictive models}
\begin{figure}[h]
  \includegraphics[width=1\columnwidth]{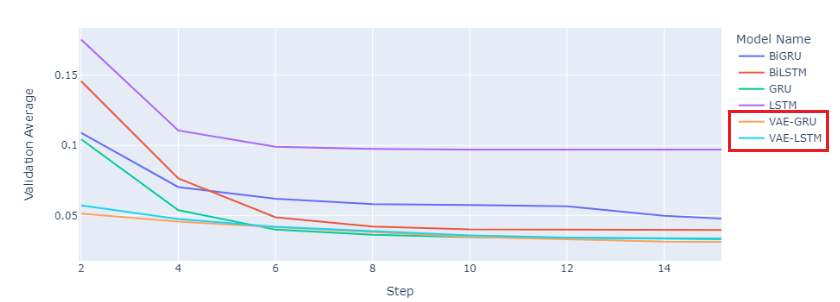}
  \caption{\textbf{Performance of different models}: The proposed model (VAE-GRU, VAE-LSTM) shows faster learning with an average of 7 epochs, while the others tend to take longer to achieve similar results.}
  \label{fig:benchmarking_performance}
\end{figure}

Beyond the accuracy of the predictive models, we assess the speed of convergence of our models compared to classical RNNs. 
We compare the proposed model in terms of resources and training time, as shown in figure~\ref{fig:benchmarking_performance}. The proposed model demonstrates \textbf{faster learning}, averaging 7 epochs to converge, while other models tend to require longer training times to achieve similar results. This efficiency makes our proposed model not only effective in clinical metrics but also resource-efficient.

\section{Conclusion and Future Works}
\label{sec:conclusion}

We introduced a novel model that reduces dependency on extensive pre-processing by leveraging the VAE latent space for data imputation and recurrent VAE for preserving temporal dynamics. 

Our proposed VAE-RNN model demonstrated superior performance in handling data preprocessing compared to manual techniques and predicting glucose levels compared to traditional RNN-based models (LSTM, Bi-LSTM, GRU, Bi-GRU). Experimental results showed that VAE-GRU achieved the lowest RMSE and MAPE values for both 30-minute and 1-hour prediction horizons. Clinically, VAE-GRU exhibited the highest accuracy in zones A and B while minimizing critical prediction errors in zones D and E. Despite the competitive performance of Bi-RNN models in quantitative metrics, they were less effective in clinical applications. Additionally, our proposed models showed significant efficiency in training time, converging in fewer epochs compared to other models.

\subsection*{Future Works}
Future research will focus on several key areas to further enhance the performance and applicability of our models:

\begin{itemize}
    \item \textbf{Model Generalization:} Exploring the generalizability of the proposed models across diverse patient populations and varying clinical settings to ensure robust performance.
    \item \textbf{Real-time Implementation:} Developing and testing real-time implementations of the models in clinical environments to evaluate their practical utility and integration into existing healthcare systems.
    \item \textbf{Multi-modal Data Integration:} Incorporating additional data sources, such as activity trackers, and dietary logs, to improve prediction accuracy and provide comprehensive patient monitoring.
    \item \textbf{Adaptive Learning:} Investigating adaptive learning techniques that allow the models to continuously learn and update from new patient data, enhancing their long-term accuracy and reliability.
    \item \textbf{Explainability and Interpretability:} Enhancing the explainability and interpretability of the models to provide healthcare professionals with actionable insights and increase trust in AI-driven decisions.
\end{itemize}


\bibliographystyle{IEEEtran}
\bibliography{main}


\end{document}